\documentclass[oneside,
twocolumn,10pt,
]{article}
\usepackage[left=2.0cm, right=2.0cm, top=2.0cm, bottom=2.0cm]{geometry}

\usepackage[utf8]{inputenc}
\usepackage[T1]{fontenc}

\usepackage{fancyhdr}
\usepackage{inconsolata}
\usepackage{mathpazo}
\usepackage[medium]{inter}

\usepackage{sectsty}
\allsectionsfont{\bfseries\sffamily}

\usepackage{titlesec}

\titlespacing*{\section}
  {0pt}   
  {2ex}   
  {1ex}   

\titlespacing*{\subsection}
  {0pt}
  {1.5ex}
  {0.8ex}

\usepackage{authblk}
\usepackage{doi}
\usepackage{svg}
\usepackage{orcidlink}
\usepackage{xcolor}
\usepackage{url}
\usepackage[font=small]{caption} 

\usepackage[
  backend=biber,
  style=phys,
  sorting=none,
  maxnames=10
]{biblatex}
\usepackage{graphicx}

\usepackage{xurl}
\usepackage{hyperref}
\hypersetup{breaklinks=true}
\usepackage{amsmath}
\usepackage{cleveref}
\usepackage{xcolor}
\definecolor{linkblue}{RGB}{0,0,120}

\hypersetup{
  colorlinks=true,
  linkcolor=linkblue,
  citecolor=linkblue,
  urlcolor=linkblue,
  breaklinks=true,
}

\newcommand{\om}{\texttt{optimade-maker}}

\title{
\om: Automated generation of interoperable materials APIs from static datasets
}

\author[1, 2, $\star$]{Kristjan Eimre \orcidlink{0000-0002-3444-3286}}
\author[3, 4, 5, $\star$, $\dagger$, $\ddagger$]{Matthew L. Evans \orcidlink{0000-0002-1182-9098}}
\author[1]{Bud Macaulay \orcidlink{0009-0009-0653-3741}}
\author[1, 2]{Xing Wang \orcidlink{0000-0003-3356-9108}}
\author[1, 2]{Jusong Yu}
\author[1, 2, 6]{Nicola Marzari \orcidlink{0000-0002-9764-0199}}
\author[3]{Gian-Marco Rignanese \orcidlink{0000-0002-1422-1205}}
\author[1, 2, $\S$]{Giovanni Pizzi \orcidlink{0000-0002-3583-4377}}

\affil[1]{PSI Center for Scientific Computing, Theory and Data, 5232 Villigen PSI, Switzerland}
\affil[2]{National Centre for Computational Design and Discovery of Novel Materials (MARVEL), 5232 Villigen PSI, Switzerland}
\affil[3]{UCLouvain, Institute of Condensed Matter and Nanosciences, Chemin des \`{E}toiles 8, Louvain-la-Neuve 1348, Belgium}
\affil[4]{Matgenix SRL, A6K Advanced Engineering Center, Charleroi, Belgium}
\affil[5]{datalab industries ltd., King's Lynn, United Kingdom}
\affil[6]{Theory and Simulation of Materials (THEOS), École Polytechnique Fédérale
de Lausanne, 1015 Lausanne, Switzerland}
\affil[$\star$]{These authors contributed equally.}
\affil[$\ddagger$]{Present address: Yusuf Hamied Department of Chemistry, University of Cambridge, Cambridge, CB2 1EW, United Kingdom}
\affil[$\dagger$]{me388@cam.ac.uk}
\affil[$\S$]{giovanni.pizzi@psi.ch}

\date{\vspace{-12pt}(\small\today)\vspace{-8pt}}

\begin{document}

\maketitle

\begin{abstract}
Atomistic structural data are central to materials science, condensed matter physics, and chemistry, and are increasingly digitised across diverse repositories and databases.
Interoperable access to these heterogeneous data sources enables reusable clients and tools, and is essential for cross-database analyses and data-driven materials discovery.
Toward this aim, the OPTIMADE (Open Databases Integration for Materials Design) specification defines a standard REST API for atomistic structures and related properties.
However, deploying and maintaining compliant services remains technically demanding and poses a significant barrier for many data providers.
Here, we present \om{}, a lightweight toolkit for the automated generation of OPTIMADE-compliant APIs directly from raw atomistic structure and property data.
The toolkit supports a wide range of raw datasets, enables conversion to a standardised OPTIMADE data representation, and allows for rapid deployment of APIs in both local and production environments.
We further demonstrate it through an automated service on the Materials Cloud Archive, which automatically creates and publishes OPTIMADE APIs for contributed datasets, enabling immediate discoverability and interoperability.
In addition, we implement data transformation pipelines for the Cambridge Structural Database (CSD) and the Inorganic Crystal Structure Database (ICSD), enabling unified access to these curated resources through the OPTIMADE framework.
By lowering the technical barriers to interoperable data publication, \om{} represents an important step toward a scalable, FAIR materials data ecosystem integrating both community-contributed and curated databases.
\end{abstract}

\section{Introduction}


Atomistic structural data of crystalline and molecular systems underpin the most fundamental aspects of materials science, condensed matter physics and chemistry.
In the digital era, such data, together with associated properties and derived quantities, are increasingly collected in structured databases and made accessible to the research community through application programming interfaces (APIs) and graphical user interfaces (GUIs) or, most often, as static datasets described by files.
These databases may contain experimentally determined structures, such as those in the Crystallography Open Database (COD) \cite{grazulis_crystallography_2012}, the Inorganic Crystal Structure Database (ICSD) \cite{zagorac_recent_2019}, the Cambridge Structural Database (CSD) \cite{Groom2016}, or the Materials Platform for Data Science (MPDS), as well as structures generated through high-throughput computational workflows \cite{horton_accelerated_2025,schmidt_dataset_2022, armiento_database-driven_2020, draxl_nomad_2018,haastrup_computational_2018, saal_materials_2013, esters_afloworg_2023, huber_mc3d_2025, mounet_two-dimensional_2018, talirz_materials_2020, Evans2020b}.



The primary APIs of atomistic structure databases, where present, are often custom-built and lack any interoperability between different providers.
Consequently, client applications aiming to access multiple databases must implement support for a variety of incompatible API standards, significantly increasing complexity for the development of clients and of data-driven pipelines that explore the data.
To address this issue, the Open Databases Integration for Materials Design (OPTIMADE) consortium, comprising representatives from many major materials databases worldwide, has developed a common specification for a REST (representational state transfer) API \cite{andersen_optimade_2021} to serve atomistic structure data and related properties.
The OPTIMADE specification is designed to accommodate the diverse requirements and constraints of materials databases, enabling uniform access to atomistic structure data across providers.
In addition, the specification defines a standard mechanism for dataset discoverability on the web, allowing compliant APIs to align with the FAIR (Findable, Accessible, Interoperable, and Reusable) data principles \cite{wilkinsonFAIRGuidingPrinciples2016}.

As a result, an OPTIMADE ecosystem has emerged.
As of March 2026, 20 database providers expose OPTIMADE-compliant APIs, collectively indexing over 25 million materials \cite{noauthor_optimade_nodate}. A growing number of clients and applications leverage the specification to discover, analyse, and aggregate materials data across multiple sources.
OPTIMADE APIs have already been successfully employed in several materials discovery and design projects \cite{evans_developments_2024, Trinquet2025, Trinquet2025a}.


Despite these advances, deploying and maintaining a materials database together with a fully compliant OPTIMADE API typically requires dedicated infrastructure consisting of hardware, software, and personnel, which is costly and time-consuming to build and maintain.
The associated technical overhead and maintenance effort present a substantial barrier for individual researchers or small research groups who wish to disseminate their data in an interoperable manner.

\begin{figure}
    \centering
    \includegraphics[width=1.0\linewidth]{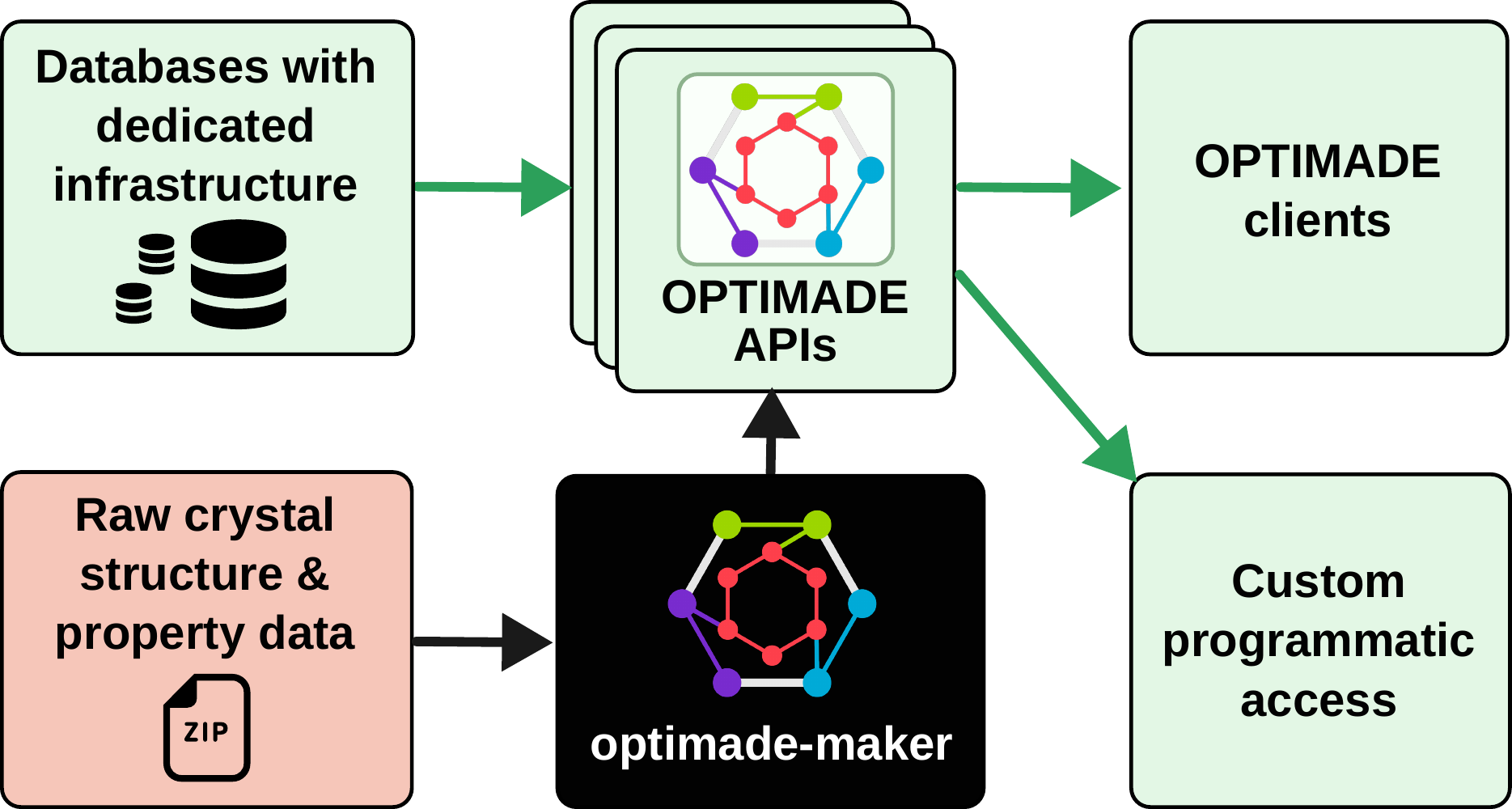}
    \caption{Schematic illustrating the context of the \om{} toolkit within the OPTIMADE ecosystem. Green boxes indicate already established entities. The red box highlights raw materials data that are not readily integrable into the ecosystem, a gap addressed by \om{}.}
    \label{fig:context}
\end{figure}

In this work, we present \om{}, a toolkit that addresses this challenge by enabling the automated generation of OPTIMADE APIs directly from raw materials data files, such as simulation outputs or structural assignments.
Built on top of the existing \texttt{optimade-python-tools} \cite{evans_optimade-python-tools_2021} Python library, \om{} can be integrated into data pipelines to provide OPTIMADE-compliant services for production web platforms, while also allowing researchers to quickly deploy a local API for using OPTIMADE-compliant clients with their raw data.
\Cref{fig:context} illustrates the position of \om{} within the OPTIMADE ecosystem.

We further present representative use cases of \om{} developed as part of this work.
One such use case is the Materials Cloud Archive OPTIMADE service, which enables researchers to automatically obtain an OPTIMADE API for their datasets upon upload to the Materials Cloud Archive \cite{talirz_materials_2020}, an open research data repository, thereby facilitating immediate interoperability and discoverability.
The Materials Cloud Archive links these datasets directly to the newly redesigned Materials Cloud OPTIMADE Client, where they can be interactively browsed.
This service has already been used to serve several contributed datasets.
Furthermore, we used \om{} to implement the OPTIMADE data transformation pipelines for the Cambridge Structural Database (CSD) and the Inorganic Crystal Structure Database (ICSD).


\section{Results} 

\subsection{\om{} toolkit}

The \om{} toolkit is developed as a Python package which enables the automated creation of OPTIMADE APIs from a range of structural data formats and associated material properties.
The toolkit can be used as a Python library or via the \texttt{optimake} command line interface (CLI) tool.
The primary features provided by the toolkit include:
1) specification of a simple YAML (YAML Ain’t Markup Language) configuration file that describes the raw data and makes it parsable for the toolkit;
2) conversion of the raw data into the standard OPTIMADE JSON Lines file format, based on JavaScript Object Notation (JSON), which was developed as part of this work and is now part of the official specification since v1.3.0;
3) serving an OPTIMADE API directly from a raw data archive or from an OPTIMADE JSON Lines file using the reference server implementation from \texttt{optimade-python-tools} \cite{evans_optimade-python-tools_2021}.


In a typical \om{} workflow, the user provides a collection of raw files describing atomistic structures, possibly in an archive file (e.g., a ZIP file), and optionally, any associated properties for these.
\om{} assigns each structure a unique identifier based on its path and file name (see \cref{sec:ids}), and the property files can reference either the identifier or the full path.
To make this data parsable by \om{}, an \texttt{optimade.yaml} configuration file is provided, describing the locations of files, and defining the relevant property metadata.
After this setup, the CLI can be used to convert the data into a standardised OPTIMADE representation and start the API.
A step-by-step, beginner-friendly tutorial demonstrating this workflow is available in the project repository, providing a guided introduction for new users.

\begin{figure}
    \centering
    \includegraphics[width=1.0\linewidth]{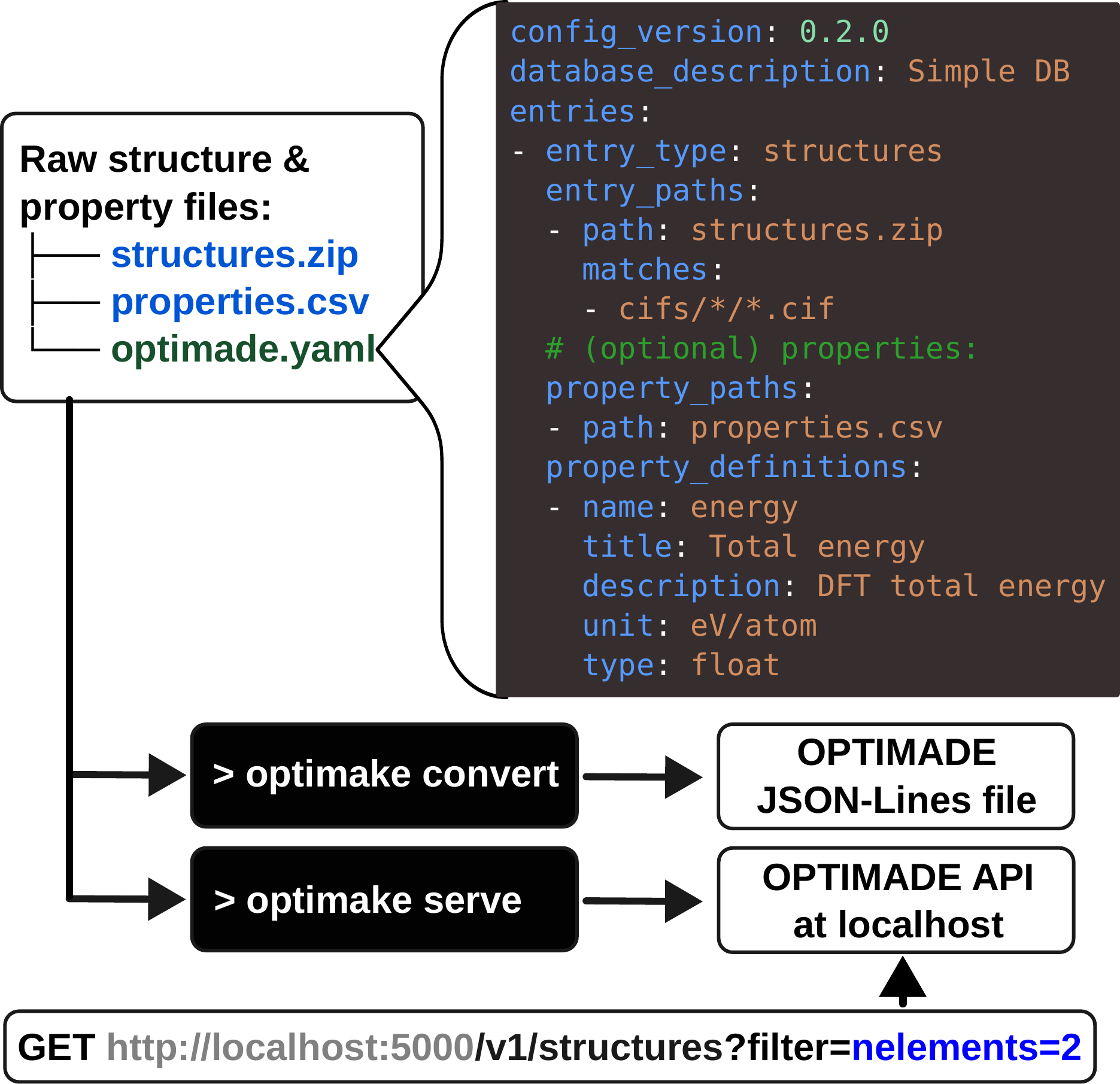}
    \caption{Schematic overview of the main components of the \om{} CLI. Raw data files are supplemented by the \texttt{optimade.yaml} configuration file, describing file locations and property definitions. Black boxes show the two primary CLI commands: \texttt{convert} transforms the raw data into the standard JSON Lines format, while \texttt{serve} launches an OPTIMADE API server. The server can be queried by any standard OPTIMADE HTTP requests, as shown in the example at the bottom (gray segment denotes the base URL, and blue segment represents a filter selecting binary structures).}
    \label{fig:optimake}
\end{figure}

A schematic overview of a concrete use case is shown in \cref{fig:optimake}.
In this example, the structures are packaged in a ZIP archive (\texttt{structures.zip}) containing multiple Crystallographic Information Files (CIFs) \cite{hallCrystallographicInformationFile1991,bernsteinSpecificationCrystallographicInformation2016}, together with a Comma-Separated Values (CSV) file (\texttt{properties.csv}) that includes, for each structure, an identifier and a property value -- the floating-point total energy per atom.
Using this configuration, the \texttt{optimake convert} CLI command can convert the raw data into the standard OPTIMADE JSON Lines format, e.g., for archival purposes.
The \texttt{optimake serve} command directly serves an OPTIMADE API from the raw data.
By default, the \texttt{serve} command starts the API locally (on \texttt{localhost}), where it is immediately available for OPTIMADE-compliant queries by client applications allowing for search across any of the standardised OPTIMADE fields and any extra properties defined in the YAML configuration file.

\om{} supports raw data in a variety of formats.
Atomistic structure files are parsed using the Atomic Simulation Environment (ASE) \cite{larsen_atomic_2017}, enabling support for most standard and non-standard formats, including CIF, XYZ, and XSF.
Pymatgen \cite{ong_python_2013} JSON files containing structures and related properties are also supported.
These files may be compressed or archived using common formats such as \texttt{.zip}, \texttt{.tar.gz}, and \texttt{.tar.bz2}.
For property data, CSV and JSON files are supported (see also \cref{sec:ids} on how to map properties to structures).

In addition, \om{} can ingest data from AiiDA \cite{pizzi_aiida_2016, huber_aiida_2020, uhrin_workflows_2021} archive files or directly from an AiiDA profile, as described in detail in \cref{sec:aiida}.


The \texttt{optimake serve} command is designed to support both rapid local deployment and more complex data pipeline setups, including production-grade APIs.
By default, the command converts the raw data and populates a temporary in-memory MongoDB database using the \texttt{MongoMock} Python library, eliminating the need for an external database and enabling immediate access to the data through an OPTIMADE API.

Alternatively, \texttt{optimake serve} can be configured using a custom \texttt{optimade-python-tools} configuration file.
This allows the population of an external MongoDB instance (e.g., for production deployments), customisation of provider metadata, or execution of the data population pipeline without starting the API itself.
The latter mode is particularly useful in automated workflows where the API is launched through separate orchestration mechanisms, e.g., as discussed in the next section.


\subsection{Materials Cloud Archive automatic OPTIMADE service}
\label{sec:mc-archive-service}

\begin{figure}
    \centering
    \includegraphics[width=1.0\linewidth]{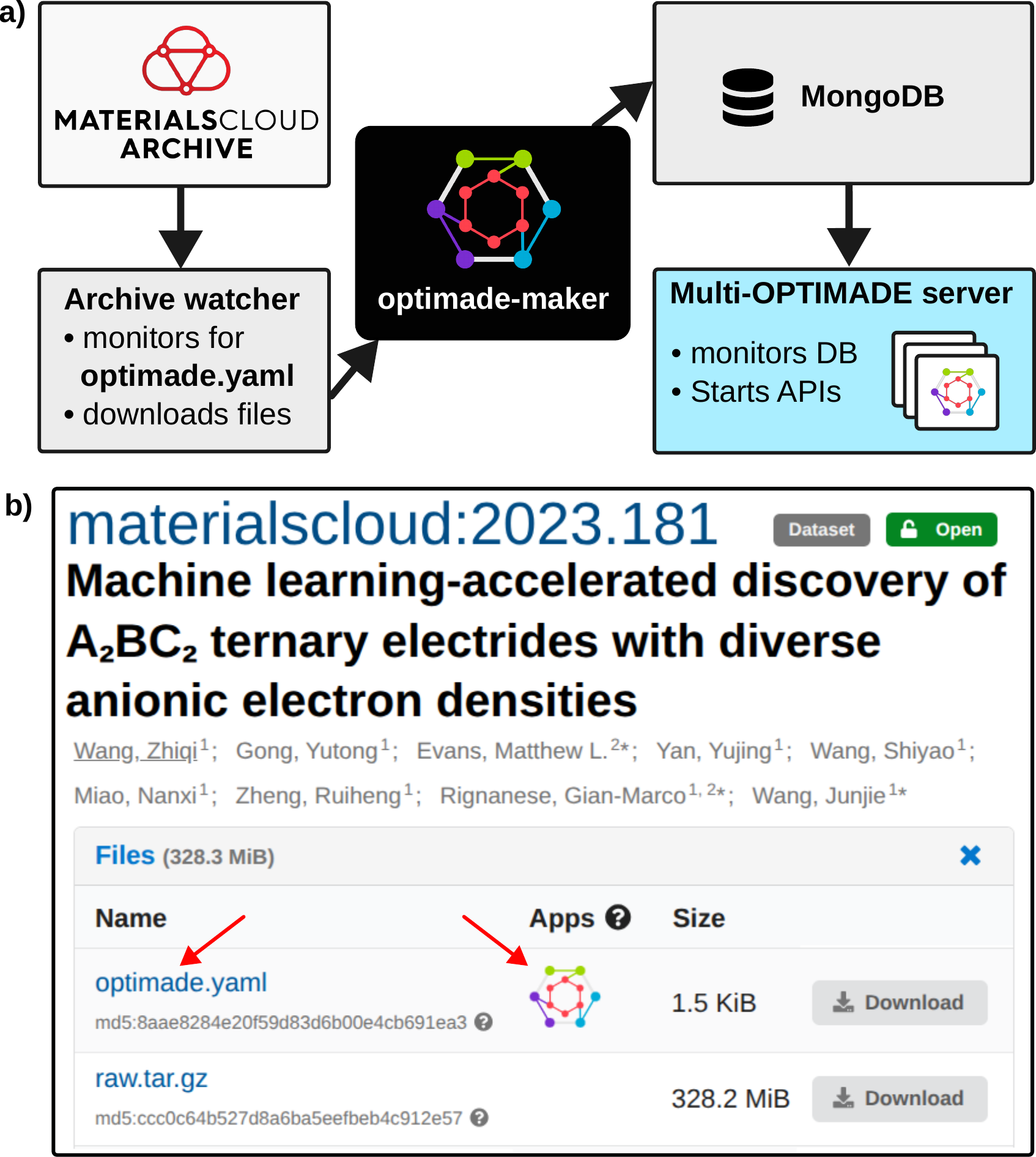}
    \caption{Materials Cloud Archive OPTIMADE service. (a) Data pipeline utilizing the \om{} toolkit. (b) A webpage for a Materials Cloud Archive entry \cite{wang_machine_2023} containing an \texttt{optimade.yaml} file, where a link is displayed to directly explore the dataset with the new Materials Cloud OPTIMADE client.}
    \label{fig:mc-archive-service}
\end{figure}

The Materials Cloud Archive \cite{talirz_materials_2020} is an open research data repository for computational materials science.
Researchers can upload datasets without restrictions on file format and make them available to the community for direct download.

Using \om{}, we developed an automated service on the Materials Cloud platform that deploys OPTIMADE APIs for archive entries compatible with the toolkit, as illustrated in \cref{fig:mc-archive-service}.
\Cref{fig:mc-archive-service}a shows a schematic overview of the data pipeline underlying this service.
A Python job, referred to as the \emph{Archive watcher}, regularly monitors newly published entries in the Materials Cloud Archive, and checks their compatibility with \om{} based on the existence of an \texttt{optimade.yaml} file in the appropriate format.
Compatible entries are processed using the \texttt{optimake serve -{}-prepare\_only} command, which converts the raw data into the required internal format and generates a corresponding \texttt{optimade-python-tools} configuration.
The processed data are then injected into a MongoDB instance, which is monitored by the \textit{Multi-OPTIMADE server}, a light wrapper around \texttt{optimade-python-tools} that allows the server to efficiently manage multiple OPTIMADE APIs.
Upon detecting new data, the \textit{Multi-OPTIMADE server} launches the corresponding APIs and mounts them under distinct subpaths of a single Python REST API.

The resulting OPTIMADE API endpoints are published to the wider OPTIMADE ecosystem under the Materials Cloud Archive provider identifier \texttt{mcloudarchive}, and can be accessed by any OPTIMADE-compatible client.
Each OPTIMADE API gets its own database identifier based on the Materials Cloud Archive DOI that represents all versions of the entry.
If a new version of an Archive entry is published, a new OPTIMADE API is set up with the same identifier that replaces the old version of the API.

The Materials Cloud Archive webpage containing the OPTIMADE configuration (\cref{fig:mc-archive-service}b) will contain a direct link to the Materials Cloud OPTIMADE Client, fully redesigned as part of this work (see details in \cref{sec:optimade-client}).
Finally, the created OPTIMADE API endpoint with all relevant metadata and links is also published to the Materials Cloud OPTIMADE overview page at \mbox{\url{https://optimade.materialscloud.org}}.

\subsection{Providing API access to materials design datasets}

The Materials Cloud Archive integration described above has been used to disseminate structures and properties for several materials discovery and design projects~\cite{evans_developments_2024}.
For instance, in Ref.~\cite{wang_machine_2023}, Wang \emph{et al.} performed high-throughput first principles calculations and trained machine learning (ML) models to screen $P\text{4}/mbm$-$A_{2}BC_{2}$ structural prototypes to design new ternary electride materials. Starting from a library of 214 known $A_2 BC_2$ phases, density-functional theory calculations were performed to assess their electride nature (via the maximum value of the electron localisation function, $\text{ELF}_\text{max}$) and to create a training set for a series of ML models. The $P\text{4}/mbm$-$A_2 BC_2$ prototype structure was then decorated with different elements to form a design space of over 14,000 hypothetical compounds and the trained ML models were used to rapidly predict their $\text{ELF}_\text{max}$ values and thermodynamic stability, identifying high priority materials to investigate further with DFT calculations. Through this approach, 41 stable and 104 metastable $A_2 BC_2$ potential electrides were predicted, with diverse anionic electron densities across the range of electron-deficient, neutral and electron-rich electrides. The three most promising materials were experimentally validated for synthesisability and catalytic activity for ammonia synthesis. The raw data from this work were deposited on the Materials Cloud Archive with a supplementary \texttt{optimade.yaml} configuration file that indicated which aspects of the dataset to index as an OPTIMADE API \cite{Wang2023_mca}; the authors chose to index the 145 compounds that were computed with DFT and their associated stabilities (formation energies, hull distances versus known phases) and their $\text{ELF}_\text{max}$ values, exposing them for further search.

In two other related projects, Trinquet \emph{et al.} performed a combined ML and DFT active learning screening for materials that exhibit strong optical responses: high refractive indices \cite{Trinquet2025} and strong second-harmonic generation for a given band gap \cite{Trinquet2025a}. 
The initial screening dataset was defined by an OPTIMADE query to curate all hypothetically stable structures that were non-centrosymmetric across contributing databases.
Then, seeding the active learning with DFT calculations on the subset of known high performing materials and a random exploratory selection of unknown materials, models were trained to predict optical response.
These models were then used to prioritise the next structures on which to perform the high-fidelity DFT calculations, with the overall search campaign comprising several such repeating loops.
The resulting datasets from each study, comprising structures, their corresponding DFT-computed refractive indices \cite{trinquet_2024_yx9a2-60n74} and their second-harmonic generation coefficients \cite{trinquet_2025_zn7cy-7cs67}, were collated and uploaded to Materials Cloud Archive.
Both datasets provide annotations describing the additional computed fields, e.g., \texttt{\_mcloudarchive\_d\_kp\_conv\_neum} with description "The effective Kurtz--Perry powder coefficient from the conventional HSE scissor-corrected SHG tensor" and unit "pm/V", which are reported in the \texttt{/info/structures} endpoint of the corresponding OPTIMADE API and are exposed for search.

Each of the three datasets described above return results when OPTIMADE users make cross-provider queries, making use of the OPTIMADE APIs served by Materials Cloud Archive.
Custom fields specified by the user are prepended with the \texttt{\_mcloudarchive} provider prefix and the underlying properties are made queryable via OPTIMADE; for example, the query \texttt{"\_mcloudarchive\_convex\_hull\_distance < 0.025 AND \_mcloudarchive\_elf\_max > 0.5"} will return all low-lying hypothetical structures from Ref. \cite{wang_machine_2023}.

\subsection{Data transformation pipelines for the CSD and ICSD}

In addition to providing API access to archived data, the tools developed in \texttt{optimade-maker} have been used to provide API access to rolling snapshots of continuously updated databases.
Two such live databases are the Cambridge Structural Database (CSD) \cite{Groom2016} and the Inorganic Crystal Structure Database (ICSD) \cite{Bergerhoff1983, zagorac_recent_2019}.
The CSD contains a "complete record of all published organic and metal-organic small molecule crystal-structures" \cite{Groom2016}, curated by experts and deposited either directly or via journal publication.
The ICSD contains "a near exhaustive list of known inorganic crystal structures published since 1913" \cite{zagorac_recent_2019}, primarily derived from diffraction data but increasingly including published theoretical structures.
As pioneers of data-driven science, both databases have been curated since the 1970s and now contain approximately 1.4 million and 327,000 entries, respectively.
However, as commercial databases, programmatic access is limited to license holders and requires bespoke software.
OPTIMADE makes no requirement that data conforming to it be open or freely available; this benefits users and database providers as tools written to target commercial (or otherwise closed) datasets that conform to OPTIMADE should also automatically work on open datasets, and vice versa, preventing further fragmentation of the ecosystem.

The UK's Physical Sciences Data Infrastructure (PSDI) \cite{PSDI_URL}, following the former Physical Sciences Data Service, provides access to both the CSD and ICSD to UK academics through a combined license.
PSDI identified the need for materials API standardisation to enable cross-search of data resources that they collect, aggregate and curate and decided upon OPTIMADE as the enabling technology.

To enable this, data pipelines were developed to map all entries in the CSD and ICSD into the OPTIMADE format, encompassing structural, bibliographic and chemical data pertaining to each entry, via the CSD Python API \cite{Sykes2024} and the ICSD REST API, respectively.
These mapped entries were then written to a combined OPTIMADE JSON Lines file which can be served as an OPTIMADE API using the tools provided by \texttt{optimade-maker}.

This approach is significantly simpler than the alternative of mapping database queries and outputs from the existing database-specific formats and returning them in an OPTIMADE compliant way, but comes at the cost of needing to run a secondary database and API that must be periodically updated from the live source.
Each database required its own extension fields to the core OPTIMADE structure type; the CSD focusing on chemical identifiers (SMILES, InChI etc.) and molecular crystal properties ($Z$, $Z'$), whereas the ICSD made use of several tabulated CIF \cite{bernsteinSpecificationCrystallographicInformation2016} fields that describe diffraction experiments such as measurement conditions and goodness-of-fit.

Although not the focus of this work, software implementations for these pipelines can be found on GitHub at \href{https://github.com/datalab-industries/csd-optimade}{datalab-industries/csd-optimade} and \href{https://github.com/datalab-industries/icsd-optimade}{datalab-industries/icsd-optimade}.
The resulting OPTIMADE APIs are not publicly accessible, as only licence holders are allowed to query and retrieve the underlying data.
However, these pipelines are deployed by the PSDI so that UK academic users can receive seamless access to structures from the CSD and ICSD in the PSDI Cross Data Search service \cite{PSDI_URL}, alongside many other resources, with unified querying and semantics to access property definitions powered by OPTIMADE.

\section{Methods}

\subsection{Software design}
\label{sec:arch}

The \om{} Python package has four components that implement the main functionality: 1) \texttt{optimade\_maker.config}, 2) \texttt{optimade\_maker.convert}, 3) \texttt{optimade\_maker.parsers}, and 4) \texttt{optimade\_maker.serve}.


\texttt{optimade\_maker.config} defines the \texttt{optimade.yaml} configuration format, making use of Pydantic \cite{pydantic_github} to provide typed, versioned schema definitions for each field. 
The configuration can be provided as JSON, YAML, or directly as a Python dictionary.
As shown in Figure \ref{fig:optimake}, the configuration consists of a few top-level metadata fields about the database itself, and then subconfiguration objects per entry type that define the files to parse as entries and, optionally, properties.
When user-defined properties are provided, they must be accompanied by extra metadata in the OPTIMADE property definition, including the field name, title, OPTIMADE data type (e.g., float or integer), unit, and a human-readable description.


\texttt{optimade\_maker.convert} implements the pipeline that takes the archived data and applies the scheme defined in the user-supplied \texttt{optimade.yaml} configuration file to create a single combined OPTIMADE JSON Lines file for the dataset.
The basic process is as follows: first, decompress the archived data (typically provided as a ZIP or tar file), then loop through the OPTIMADE resource types (e.g., structures, references) that have processing rules provided in the configuration file. 
These rules include a list of patterns that match file paths to attempt to parse as the given entry type, \texttt{entry\_paths} (e.g., the wildcard \texttt{*.cif}), and an optional list of \texttt{property\_paths} corresponding to auxiliary files that contain property data pertaining to those entries.
The properties themselves need to be defined in the \texttt{property\_definitions} field of the configuration.
The conversion process then constructs the appropriate entries for each entry type, decorates them with any provided properties, creates the corresponding \texttt{/info/<type>} resource that describes the user-extended entry type in the OPTIMADE format, and finally saves each entry as an individual line in a combined JSON Lines file.


The \texttt{optimade\_maker.parsers} module is a registry of tools that map files of a given representation (standardised or otherwise) into intermediate objects that can be mapped to OPTIMADE entries. 
These tools tend to be implemented in other open source libraries, such as ASE \cite{larsen_atomic_2017} or pymatgen \cite{ong_python_2013}.
These libraries, combined, provide parsers for many atomistic simulation codes and otherwise standardised structural file formats (e.g., CIF, XYZ), with light wrappers to allow the parsers to fail fast.
For structural data, parsers can return either ASE \texttt{Atoms} or Pymatgen \texttt{Structure} objects, which are then mapped to OPTIMADE structures using the adapters implemented in \texttt{optimade-python-tools} \cite{evans_optimade-python-tools_2021}.
Rather than relying on file extensions, each parser is run in turn on each file until it can be successfully parsed and converted into an OPTIMADE structure.
A similar process is followed for parsing property data, although here only CSV and JSON files are supported, which can both be unambiguously read using the \texttt{pandas} library or the Python standard library.
Extra validation is performed against the user-provided property definitions, ensuring that data types can be appropriately coerced and IDs can be uniquely matched to the created entries (see Section \ref{sec:ids}).
These parser components can be easily extended to accommodate new libraries, file formats and entry types.


Finally, \texttt{optimade\_maker.serve} takes the constructed OPTIMADE JSON Lines file and serves it with an OPTIMADE API using the reference server implementation from \texttt{optimade-python-tools} (based on FastAPI and leveraging MongoDB as the database backend), which supports the majority of OPTIMADE 1.3 features and enables search over both standard and user-provided properties.
This allows future updates (e.g., new features, performance improvements) to \texttt{optimade-python-tools} and the OPTIMADE specification itself to be readily accommodated in \om{}.

\subsection{Structure identifiers}
\label{sec:ids}

As each structure must have an identifier in OPTIMADE, \om{} generates, by default, a structure identifier (\texttt{id}) based on its path relative to the \texttt{optimade.yaml} configuration file.
The identifier is constructed using a simple deterministic rule: from the set of all file paths, the longest common prefix and suffix (including file extensions) shared by all paths are removed.
For example, given the two structures: \texttt{structures.zip/cifs/set1/101.cif} and \texttt{structures.zip/cifs/set2/102.cif}, the corresponding identifiers are \texttt{set1/101} and \texttt{set2/102}.

\subsection{AiiDA integration}
\label{sec:aiida}

We also implement the ability for \om{} to create an OPTIMADE API directly from AiiDA \cite{pizzi_aiida_2016, huber_aiida_2020, uhrin_workflows_2021} databases.
The AiiDA workflow management infrastructure allows users to define and execute computational workflows, automatically storing the results and their full provenance in a graph database where each data object and process is represented as a node.

The \om{} package can access either a live AiiDA database or an exported \texttt{.aiida} archive file.
The user should specify an AiiDA group containing the subset of the structures to be exposed through the OPTIMADE API.
AiiDA UUIDs (universally unique identifiers) are used for the OPTIMADE structure identifiers.
Additional properties can be associated with each structure by extracting them either directly from the AiiDA structure node (e.g., from the extras) or from other nodes, according to a query encoded in the property definition.

\begin{figure}
    \centering
    \includegraphics[width=1.0\linewidth]{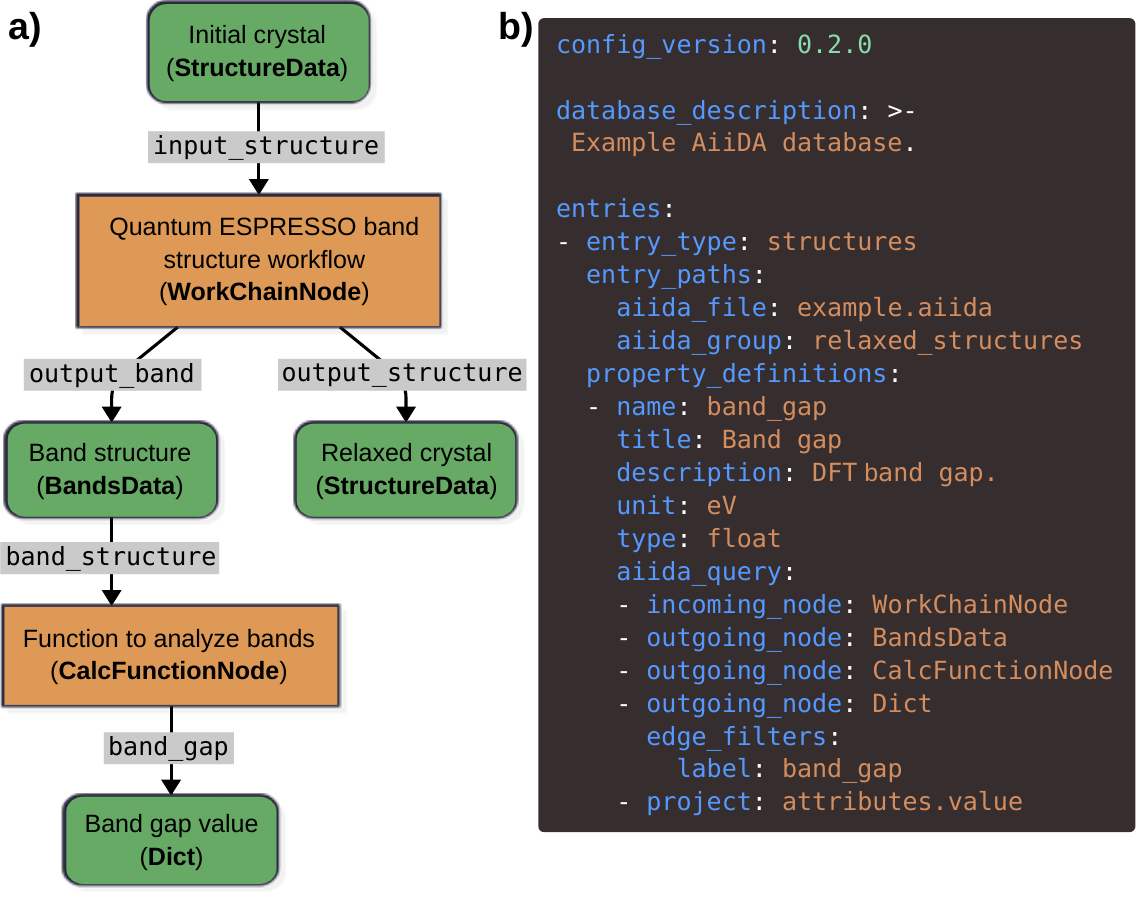}
    \caption{\om{} integration with AiiDA. (a) A schematic representing an AiiDA provenance graph. Green rounded rectangles and orange rectangles represent data and process nodes, respectively. Each node contains a description, and its AiiDA type (bold). Labels on the arrows represent AiiDA edge labels. (b) The \om{} configuration file that allows to convert the AiiDA database into the OPTIMADE format, and serve it via the API.}
    \label{fig:aiida}
\end{figure}

\Cref{fig:aiida}a shows a schematic example of an AiiDA provenance graph representing a density-functional theory (DFT) crystal-structure relaxation followed by an electronic band-structure calculation, using the Quantum ESPRESSO \cite{giannozziQUANTUMESPRESSOModular2009, giannozzi_advanced_2017} DFT code.
The workflow internally performs the two steps, producing data nodes as outputs.
The band-structure node does not directly store the band-gap value; instead, this value is computed in an additional postprocessing step, which returns a dictionary (\texttt{Dict}) node.
To use \om{} to create an OPTIMADE database and API for the relaxed structures in this AiiDA database, and also associate the calculated band-gap values as properties of each structure, the user can prepare the configuration file shown in \cref{fig:aiida}b.
Structures are selected from the AiiDA group \texttt{relaxed\_structures}.
The property definition for the band gap includes an \texttt{aiida\_query} section, which defines a query for the relevant property node relative to the structure node, using the standard filtering and projection keywords provided by the AiiDA \texttt{QueryBuilder}.

Finally, the functionality supporting AiiDA databases is lazily loaded and is not a mandatory dependency, so that \om{} does not require an AiiDA installation for its basic functionality.


\subsection{New Materials Cloud OPTIMADE Client}
\label{sec:optimade-client}

\begin{figure}
    \centering
    \includegraphics[width=1.0\linewidth]{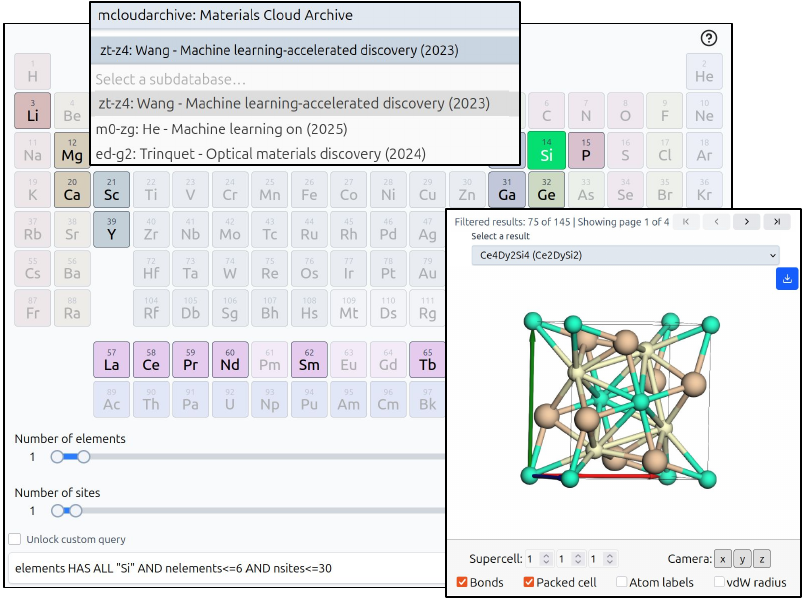}
    \caption{Screenshot of the new Materials Cloud OPTIMADE Client. An OPTIMADE provider and database (here, a contributed dataset from the Materials Cloud Archive) are selected via dropdown menus. Materials are filtered by composition using an interactive periodic table and by structural properties using sliders. The inset shows the results section -- a 3D visualisation of a filtered structure.}
    \label{fig:mc-optimade-client}
\end{figure}

In addition to enabling researchers to automatically publish their data to the OPTIMADE ecosystem, the Materials Cloud platform provides the OPTIMADE Client, a web application for interactively exploring OPTIMADE-compliant datasets.
As part of this work, to enhance the automatic Materials Cloud Archive OPTIMADE service presented in \cref{sec:mc-archive-service}, the OPTIMADE Client underwent a complete overhaul, rewriting it from Python to JavaScript and resulting in substantial improvements in performance and usability.
\Cref{fig:mc-optimade-client} shows a screenshot of the updated web application.
The client automatically displays a list of all public OPTIMADE providers and their databases, including contributed entries from the Materials Cloud Archive, and allows users to select among them.
Custom OPTIMADE API URLs are also supported, enabling, for example, the exploration of locally hosted APIs started with \om{}.
To filter materials, the client provides an interactive periodic table for selecting compositions, as well as sliders for constraining structural properties.
Any selected filters are reflected in a textbox containing the corresponding raw OPTIMADE query string, which can be reused in other applications or for learning the query language.
This textbox can also be edited manually, enabling the construction of more complex OPTIMADE queries.
The resulting structures can be browsed and visualised, and any associated properties are displayed.
Selected structures can be downloaded in multiple formats or directly imported into the Materials Cloud Quantum ESPRESSO input generator web application.
The new OPTIMADE Client is publicly available at \mbox{\url{https://optimadeclient.materialscloud.io}}.

\section{Conclusions and Outlook}

In this work, we introduced \om{}, a toolkit for the automated generation of OPTIMADE-compliant APIs directly from raw materials data.
We demonstrated the flexibility and practical impact of the toolkit through its integration into the Materials Cloud ecosystem, in particular by facilitating the automatic deployment of OPTIMADE APIs for datasets contributed by the community to the Materials Cloud Archive \cite{talirz_materials_2020}. 
Furthermore, we introduced a new Materials Cloud OPTIMADE client to make any data exposed via OPTIMADE easily accessible through an intuitive web GUI.
We also described the mapping and serving of existing live databases through OPTIMADE for the Cambridge Structural Database (CSD) and the Inorganic Crystal Structure Database (ICSD).

\om{} and the automated Materials Cloud Archive OPTIMADE service mark an important step toward a fully FAIR (Findable, Accessible, Interoperable, and Reusable) ecosystem for materials science.
By automatically exposing raw datasets through a standardized API and registering them in the OPTIMADE provider registry, datasets become immediately discoverable and interoperable across the OPTIMADE ecosystem.
This enables seamless access through a wide range of clients and tools while placing no additional burden on data contributors, lowering the barrier to FAIR data publication and reuse.


Future developments in the OPTIMADE specification could further enhance this framework.
Recent additions, including new entry types such as \texttt{files} and \texttt{trajectories} and an updated property definition format with controlled vocabularies and richer array typing \cite{evans_developments_2024}, could be incorporated into \om{}.
Support for externally defined community properties may also enable richer queries over existing datasets; for example, the \texttt{cheminfo} namespace \cite{cheminfo_namespace} could enable molecular substructure searches via SMILES annotations.

More broadly, the approach could extend beyond structural data.
By leveraging the federated registry of machine-actionable parsers developed in the \emph{datatractor} initiative \cite{Evans2025}, similar API-based access could be provided for experimental datasets while reusing many of the components of \om{}.




\section*{Data availability}

The source code of \texttt{optimade-maker} is released under the permissive MIT license and is available on GitHub at \href{https://github.com/Materials-Consortia/optimade-maker/}{Materials-Consortia/optimade-maker} and archived on Zenodo at \href{https://doi.org/10.5281/zenodo.18863676}{DOI: 10.5281/zenodo.18863676}. The datasets served by the Materials Cloud Archive OPTIMADE service are available on the Materials Cloud Archive \cite{talirz_materials_2020} under Creative Commons licenses. 

\section*{Acknowledgements}

K.E., B.M., X.W., J.Y., N.M. and G.P. acknowledge funding by the NCCR MARVEL, a National Centre of Competence in Research, funded by the Swiss National Science Foundation (grant number 205602), and by the Open Research Data Program of the ETH Board (projects ``API-03 IntER'' and ``PREMISE'': Open and Reproducible Materials Science Research).
M.L.E. thanks the BEWARE scheme of the Wallonia-Brussels Federation for funding under the European Commission's Marie Curie-Skłodowska Action (COFUND 847587) and the Leverhulme Trust for funding via an Early Career Research Fellowship.
M.L.E. also thanks the UK's Physical Sciences Data Infrastructure (EP/X032701/1) and the Cambridge Crystallographic Data Centre, in particular Prof Simon Coles, Dr Ian Bruno and Dr Mehmet Giritli, for coordinating and motivating aspects of this work via commercial engagement with datalab industries ltd, and Dr Cameron Hargreaves for assistance on the ICSD pipeline.
We thank Valeria Granata for assistance with the Materials Cloud Archive integration.

\section*{Competing Interests}

M.L.E. is the founder and director of datalab industries ltd.

\section*{Author contributions}


K.E.: conceptualisation, data curation, investigation, methodology, resources, software, validation, visualisation, writing - original draft, writing - review and editing.
M.L.E.: conceptualisation, data curation, funding acquisition, investigation, methodology, resources, software, validation, writing - original draft, writing - review and editing.
B.M.: software, visualisation, writing - review and editing.
X.W.: conceptualisation, software, writing - review and editing.
J.Y.: conceptualisation, software, writing - review and editing.
N.M.: conceptualisation, funding acquisition, resources, writing - review and editing.
G-M.R.: conceptualisation, funding acquisition, resources, supervision, writing - review and editing.
G.P.: conceptualisation, funding acquisition, resources, supervision, writing - review and editing.

\printbibliography
\end{document}